\documentclass[sigconf, nonacm]{acmart}
\AtBeginDocument{%
  }

\copyrightyear{2026}
\acmYear{2026}
\acmConference[WDSFPGA'26]{2nd Workshop on Domain-Specialized FPGAs}{Feb 22, 2026}{Seaside, CA}

\begin{document}
\settopmatter{printfolios=true}

\title{Position: The Need for Ultrafast Training}
\subtitle{Position paper at the 2nd Workshop on Domain-Specialized FPGAs (WDSFPGA 2026)}

\author{Duc Hoang}
\email{dhoang@mit.edu}
\orcid{1234-5678-9012}
\affiliation{%
  \institution{Massachusetts Institute of Technology}
  \city{Cambridge}
  \state{MA}
  \country{USA}
}

\sloppy


\begin{abstract}
Domain-specialized FPGAs have delivered unprecedented performance for low-latency inference across scientific and industrial workloads, yet nearly all existing accelerators assume static models trained offline, relegating learning and adaptation to slower CPUs or GPUs.
This separation fundamentally limits systems that must operate in non-stationary, high-frequency environments, where model updates must occur at the timescale of the underlying physics.
In this paper, I argue for a shift from inference-only accelerators to ultrafast on-chip learning, in which both inference and training execute directly within the FPGA fabric under deterministic, sub-microsecond latency constraints. 
Bringing learning into the same real-time datapath as inference would enable closed-loop systems that adapt as fast as the physical processes they control, with applications spanning quantum error correction, cryogenic qubit calibration, plasma and fusion control, accelerator tuning, and autonomous scientific experiments.
Enabling such regimes requires rethinking algorithms, architectures, and toolflows jointly, but promises to transform FPGAs from static inference engines into real-time learning machines.
\end{abstract}


\begin{teaserfigure}
  \includegraphics[width=\textwidth]{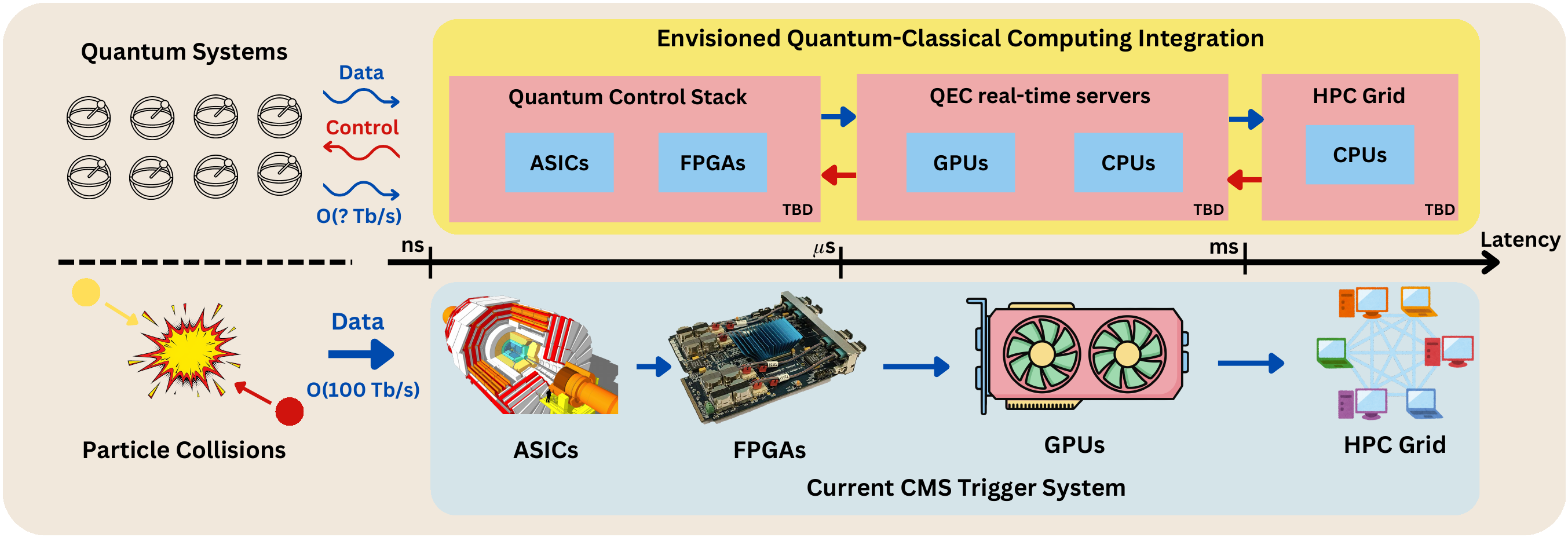}
  \caption{The missing layer in today’s FPGA accelerators.
  Hierarchical computing architectures are already proven, for example, at the LHC, where ASICs and FPGAs execute deterministic decisions within nanoseconds while higher tiers perform slower optimization.
  Emerging quantum systems demand the same structure but additionally require continuous recalibration and learning \cite{quantum_scaling}.
  This paper argues that FPGAs must evolve from inference-only accelerators into ultrafast learning engines, enabling real-time adaptation at the timescale of the underlying physics.}
  \Description{}
  \label{fig:teaser}
\end{teaserfigure}

\maketitle

\section{Introduction}

In recent years, the FPGA community has achieved remarkable success in accelerating machine learning inference under extreme latency and throughput constraints.
Frameworks such as \textsc{hls4ml}~\cite{hls4ml} and FINN~\cite{FINN} have shown that neural networks can be compiled into deeply pipelined hardware, delivering deterministic, nanosecond-scale responses for applications ranging from particle-physics triggers~\cite{cms_l1} to accelerator control~\cite{hls4ml_accelerator_control}, fusion systems~\cite{hls4ml_fusion}, and qubit readout~\cite{hls4ml_qubit_readout}.

However, nearly all of the current frameworks share a hidden assumption: \emph{learning happens elsewhere}.
Models are trained offline on GPUs or CPUs, frozen, and then deployed to FPGAs as static circuits.
FPGAs are therefore primarily used as fixed-function inference engines.

This conventional separation between learning and execution is increasingly misaligned with the systems we seek to control.
Many emerging workloads, such as quantum error correction \cite{FPGA_Surface_Code}, superconducting qubit calibration \cite{QuantumRL_Nature}, plasma confinement \cite{hls4ml_fusion}, adaptive optics\cite{adaptive_optics}, and real-time scientific instrumentation \cite{real_time_axis_control}, operate in non-stationary regimes where system dynamics evolve continuously on microsecond or even nanosecond timescales.
In these settings, a model trained even moments earlier may already be stale.
Offloading gradient computations to a host processor and returning updated parameters over high-latency interconnects exceeds the latency budget of the control loop and makes closed-loop adaptation impractical.
Therefore, learning must occur at the same temporal scale as inference within the hardware itself, or the system will inevitably lag behind the dynamics it seeks to control.

This paper takes a position: the next generation of domain-specialized fabrics should support \textbf{ultrafast on-chip learning}, in which both inference and gradient-based updates are computed directly on streaming data with deterministic, sub-microsecond latency. Rather than treating training as an offline, host-side procedure, learning must be integrated into the hardware datapath itself, operating continuously and synchronously with the physical process under control. Only by collapsing the boundary between learning and execution can adaptive systems remain stable and responsive at these timescales.

Realizing this vision demands co-design across learning algorithms, hardware architectures, and CAD toolflows, as methods developed for offline, host-driven training assume abundant memory, floating-point arithmetic, and relaxed timing, which break down under hard real-time constraints. 
Instead, in on-chip learning, updates must be structured around low-latency, fixed-precision computation under strict memory constraints.
This shift fundamentally changes what constitutes a “practical” learning system: rather than maximizing throughput or statistical efficiency in large batches, the objective becomes minimizing latency, memory footprint, and worst-case update cost.
Embracing these constraints enables a new regime of adaptive computation, in which systems learn continuously from streaming data and react at the timescale of the underlying dynamics, which is unattainable with today’s frozen, periodically retrained models.

\section{A Quantum Gap}

Quantum processors are fundamentally analog control systems.
Their behavior depends continuously on a high-dimensional set of hardware parameters such as bias voltages, pulse amplitudes, frequencies, phases, and timing skews.
Small environmental drifts and device-level fluctuations therefore translate directly into performance loss, even when the quantum circuit itself is logically digital.
As a result, utility-scaler operation requires not only quantum error correction (QEC) to suppress stochastic faults, but also continual \emph{calibration} to keep the underlying physical error rates safely below the QEC threshold \cite{RL_GoogleQuantumAI, SCALABLE_QUANTUM}.
Crucially, the relevant control parameters are non-stationary: operating points can shift over time due to slow drift, sporadic rearrangements, and measurement backaction, and different error mechanisms evolve on different timescales.
This non-stationarity turns calibration from a one-time procedure into an ongoing control problem, especially when the goal is sustained, autonomous execution under tight feedback and determinism constraints \cite{FPGA_Surface_Code,QuantumRL_Nature}.

This creates a systems-level gap between what large-scale quantum machines need and what today’s control stacks provide. Most calibration remains episodic and host-driven: measurements are acquired, shipped to CPUs/GPUs, processed with substantial software overhead, and only then used to update control settings \cite{RL_GoogleQuantumAI, QuantumRL_Nature}. At scale, this approach is brittle because drift is non-stationary, so the “right” parameters are a moving target, and because intermittent recalibration pauses are incompatible with the continuous runtimes envisioned for fault-tolerant algorithms that may need to run for days or weeks.

What fails is not learning in principle, but learning in deployment. Reinforcement learning offers a principled way to calibrate under uncertainty by optimizing control actions from streaming reward signals without an explicit differentiable model of the device \cite{QuantumRL_Nature, Sivak_2023, Baum_2021}.
However, implementing RL on a host reintroduces exactly the non-deterministic software and interconnect latencies that defeat fast stabilization.
In that sense, quantum control inherits the same hidden assumption common in today’s FPGA ML toolflows: \emph{learning happens elsewhere}.

A concrete instance of this quantum gap arises in semiconductor quantum dot platforms \cite{QUANTUM_DOT}, where charge sensors and operating points are sensitive to low-frequency $1/f$ charge noise.
In these devices, small charge rearrangements can shift Coulomb peaks and degrade charge-sensing contrast, requiring frequent re-centering of the operating point.
Today this is often handled through manual, time-intensive ``knob tuning'' or intermittent, host-driven optimization loops that interrupt operation and do not scale cleanly with increasing device complexity.
The key opportunity is to replace these slow outer loops with \emph{on-chip} learning controllers that ingest streaming measurements and apply corrective actions with deterministic and fast feedback.

This is where ultrafast on-chip learning becomes an enabling capability.
By combining RL-style closed-loop control with on-chip function approximators that are stable in fixed precision, one can run calibration and stabilization cycles at sub-microsecond latency.
The target operating point is a calibration loop as fast as $1~\mu s$, which is below a representative qubit decoherence timescale in silicon qubits ($\approx 10~\mu s$).
Operating in this regime can effectively average over slow charge noise while still reacting to higher-frequency fluctuations, potentially increasing the qubit coherence time by orders of magnitude.

In summary, the quantum gap is the mismatch between (i) continuously drifting analog hardware that demands constant, low-latency adaptation and (ii) control stacks and ML toolflows that assume offline training and infrequent parameter updates.
Closing this gap requires treating learning as a real-time primitive in the control hardware.
From the FPGA perspective, quantum calibration is the next frontier: it demands deterministic, streaming, fixed-precision learning at sub-microsecond latency, precisely the regime that today’s inference-centric frameworks do not address.

\section{Potential Impact}

Broadly speaking, realizing ultrafast on-chip learning would enable a new class of adaptive systems currently impossible with static models.

\textbf{Autonomous quantum calibration.}
Rather than slow, disruptive calibration cycles, quantum control systems could maintain optimal operating points through continuous, imperceptible adaptation.
Single-shot readout discriminators could track slow drifts in measurement chains.
Charge sensors could autonomously stabilize against $1/f$ noise.
This capability alone could extend effective coherence times by orders of magnitude.

\textbf{Self-tuning scientific instruments.}
Particle physics detectors, astronomical instruments, and medical imaging systems all require careful calibration that currently demands significant human intervention.
On-chip learning would enable instruments that continuously self-calibrate, maintaining optimal performance despite environmental changes, component aging, and configuration drift.

\textbf{Real-time control under non-stationarity.}
Plasma confinement, adaptive optics, and high-frequency trading all involve control problems where the dynamics being controlled change faster than offline retraining can track.
On-chip learning would enable control policies that adapt in real-time, tracking non-stationary dynamics rather than averaging over them.

\section{Why On-Chip Learning Is Hard}
\label{sec:why_hard}

On-chip learning imposes constraints that most ``offline'' training pipelines never face: deterministic ultrafast latency, bounded local memory, and fixed-precision arithmetic.
These constraints make learning qualitatively harder than inference, even on the same fabric.

\textbf{Hard real-time determinism.}
In closed-loop systems, occasional latency spikes are not acceptable: inference, gradient computation, and parameter updates must execute with fixed timing (or tightly bounded jitter), otherwise the controller can violate stability and safety margins \cite{real_time_axis_control}.
Host--accelerator loops add interconnect and scheduling variability that is fundamentally misaligned with these requirements \cite{RL_GoogleQuantumAI, QuantumRL_Nature}.

\textbf{Training amplifies compute and data movement.}
Compared to inference, gradient-based updates increase arithmetic intensity and introduce additional dependencies (e.g., error propagation, reductions), while also requiring activation/gradient buffering.
This expands both logic and memory bandwidth demand and makes timing closure substantially harder under sub-microsecond budgets \cite{ML_FPGA_Survey, EF_TRAIN}.

\textbf{Memory is the bottleneck, not just FLOPs.}
Storing activations, gradients, and optimizer state competes with on-chip LUTRAM/BRAM/URAM capacity and port bandwidth; access patterns become a first-order design constraint.
These costs grow quickly with model size and are difficult to amortize in streaming or small-batch regimes typical of real-time control \cite{ML_FPGA_Survey}.

\textbf{Reduced precision destabilizes optimization.}
FPGAs favor fixed-point arithmetic for latency and area, but gradient updates are numerically fragile: quantization, saturation, and limited dynamic range can cause divergence or stagnation unless scaling and update rules are carefully designed for fixed precision \cite{ML_FPGA_Survey, EF_TRAIN}.

\textbf{Toolflows are inference-centric.}
Current compiler and deployment stacks largely optimize static forward graphs for inference (e.g., \textsc{hls4ml} and FINN) \cite{hls4ml, FINN}. Stateful, continuous training complicates verification and bit-accurate reproducibility, and small numerical discrepancies can accumulate over time, which helps explain why many practical systems keep learning off-chip or episodic \cite{RL_GoogleQuantumAI}.

\section{Possible Research Direction}

This section is purely speculative. Ultrafast on-chip learning will likely require coordinated advances across learning algorithms, hardware architectures, and design automation toolflows.

\textbf{Algorithms.}
We need systematic characterization of which model classes admit efficient on-chip training.
Sparse or structured designs may reduce update cost and memory traffic.
The community should develop benchmarks and metrics that evaluate trainability under hard real-time constraints, not just inference efficiency.

\textbf{Architectures.}
FPGA vendors should consider architectural features that support online learning: efficient sparse update engines, gradient accumulation buffers, and mechanisms for parameter updates that maintain inference determinism during learning.
The distinction between ``inference fabric'' and ``training fabric'' may need to blur.

\textbf{CAD tools.}
Today’s HLS and synthesis flows are optimized for static datapaths and forward-only kernels.
On-chip learning demands CAD support for \emph{stateful} designs with tightly bounded latency: explicit update logic, persistent parameter storage, and guaranteed worst-case scheduling under streaming I/O.
Rather than “automatically inserting backprop,” the key need is tool support to co-optimize compute, memory, and control so that updates meet hard real-time timing and resource constraints.

\section{Conclusion}

FPGAs have proven they can deliver deterministic, nanosecond-scale inference, but today’s accelerators remain constrained by a default assumption that learning happens off-chip.
For emerging non-stationary systems, especially quantum control and other high-frequency scientific workloads—this separation makes closed-loop adaptation too slow and too unpredictable.

This paper argued for ultrafast on-chip learning as a next step: integrating inference, gradient computation, and parameter updates directly into the FPGA datapath under sub-microsecond latency bounds.
Achieving this will require joint progress in learning algorithms that are stable in fixed precision, architectures that make state updates efficient without sacrificing determinism, and CAD toolflows that treat continuous, stateful learning as a compilation target.
If realized, this shift would transform FPGAs from static inference engines into real-time learning machines, enabling adaptive instruments and controllers that track fast dynamics as it evolves.

\section{Acknowledgments}

D.H. is supported by the Phil Harris Research Group at MIT. The author would also like to acknowledge Jann Ungerer, Phuong Nguyen, and Julian Santen of the Yacoby Lab at Harvard for insightful discussions on quantum technologies.

\bibliographystyle{ACM-Reference-Format}
\bibliography{bib/references}

\end{document}